\documentclass[sigconf]{acmart} 
\AtBeginDocument{%
  }

\setcopyright{acmlicensed}

\copyrightyear{2026}
\acmYear{2026}
\setcopyright{cc}
\setcctype{by}
\acmConference[CHI '26]{Proceedings of the 2026 CHI Conference on Human Factors in Computing Systems}{April 13--17, 2026}{Barcelona, Spain}
\acmBooktitle{Proceedings of the 2026 CHI Conference on Human Factors in Computing Systems (CHI '26), April 13--17, 2026, Barcelona, Spain}
\acmDOI{10.1145/3772318.3791089}
\acmISBN{979-8-4007-2278-3/2026/04}

\usepackage{xcolor}
\definecolor{grey}{rgb}{0.5,0.5,0.5}

\usepackage{fontawesome5} 

\usepackage{framed}
\usepackage{xcolor}

\definecolor{sarahbg}{RGB}{255, 240, 230}
\definecolor{sarahframe}{RGB}{205, 120, 50}

\newenvironment{sarahexample}{%
    \MakeFramed{\advance\hsize-\width \FrameRestore}%
    \setlength{\parindent}{0pt}%
    \setlength{\parskip}{3pt}
    {\sffamily\small\textcolor{gray}{\faUser}\ \textbf{Sarah's Scenario:}}%
    \par  
    \small\normalfont\ignorespaces
}{%
    \endMakeFramed
}

\usepackage{listings}
\usepackage{pifont}
\usepackage{xcolor}

\definecolor{insightbg}{RGB}{250, 250, 250}
\definecolor{insightframe}{RGB}{150, 150, 150}
\definecolor{insightaccent}{RGB}{205, 100, 30}

\newenvironment{designinsightbox}{%
    \MakeFramed{\advance\hsize-\width \FrameRestore}%
    \setlength{\parindent}{0pt}%
    \setlength{\parskip}{3pt}
    \textit{Design Recommendation(s):}\par
    \small\noindent\ignorespaces
}{%
    \endMakeFramed
}

\newenvironment{insightbox}{%
    \MakeFramed{\advance\hsize-\width \FrameRestore}%
    \setlength{\parindent}{0pt}%
    \setlength{\parskip}{3pt}
    \small\noindent\ignorespaces
}{%
    \endMakeFramed
}
\definecolor{darkgreen}{RGB}{0,120,0}

\begin{document}

\author{Shreya Bali}
\email{sbali@andrew.cmu.edu}
\orcid{0000-0002-6005-409X}
\affiliation{%
  \institution{Human-Computer Interaction Institute\\Carnegie Mellon University}
  \city{Pittsburgh}
  \state{Pennsylvania}
  \country{USA}
}

\author{Riku Arakawa}
\email{rarakawa@cs.cmu.edu}
\orcid{0000-0001-7868-4754}
\affiliation{%
  \institution{Human-Computer Interaction Institute\\Carnegie Mellon University}
  \city{Pittsburgh}
  \state{Pennsylvania}
  \country{USA}
}

\author{Peace Odiase}
\orcid{0009-0007-5929-8795}
\email{odiase.peace@medstudent.pitt.edu}
\affiliation{%
  \institution{University of Pittsburgh}
  \city{Pittsburgh}
  \state{Pennsylvania}
  \country{USA}
}

\author{Tongshuang Wu}
\orcid{0000-0003-1630-0588}
\email{sherryw@cs.cmu.edu}
\affiliation{%
  \institution{Human-Computer Interaction Institute\\Carnegie Mellon University}
  \city{Pittsburgh}
  \state{Pennsylvania}
  \country{USA}
}

\author{Mayank Goel}
\orcid{0000-0003-1237-7545}
\email{mayankgoel@cmu.edu}
\affiliation{%
  \institution{School of Computer Science\\Carnegie Mellon University}
  \city{Pittsburgh}
  \state{Pennsylvania}
  \country{USA}
}

\title{Evidotes: Integrating Scientific Evidence and Anecdotes to Support Uncertainties Triggered by Peer Health Posts}


\renewcommand{\shortauthors}{Bali et al.}



\begin{abstract}
Peer health posts surface new uncertainties, such as questions and concerns for readers. Prior work focused primarily on improving relevance and accuracy fails to address users' diverse information needs and emotions triggered. Instead, we propose directly addressing these by information augmentation. We introduce Evidotes, an information support system that augments individual posts with relevant scientific and anecdotal information retrieved using three user-selectable lenses (dive deeper, focus on positivity, and big picture). In a mixed-methods study with 17 chronic illness patients, Evidotes improved self-reported information satisfaction (3.2$\rightarrow$4.6) and reduced self-reported emotional cost (3.4$\rightarrow$1.9) compared to participants' baseline browsing. Moreover, by co-presenting sources, Evidotes unlocked \textit{information symbiosis}: anecdotes made research accessible and contextual, while research helped filter and generalize peer stories. Our work enables an effective integration of scientific evidence and human anecdotes to help users better manage health uncertainty.
\end{abstract}



\maketitle

\begin{figure*}[t]
  \centering
  \includegraphics[width=\textwidth]{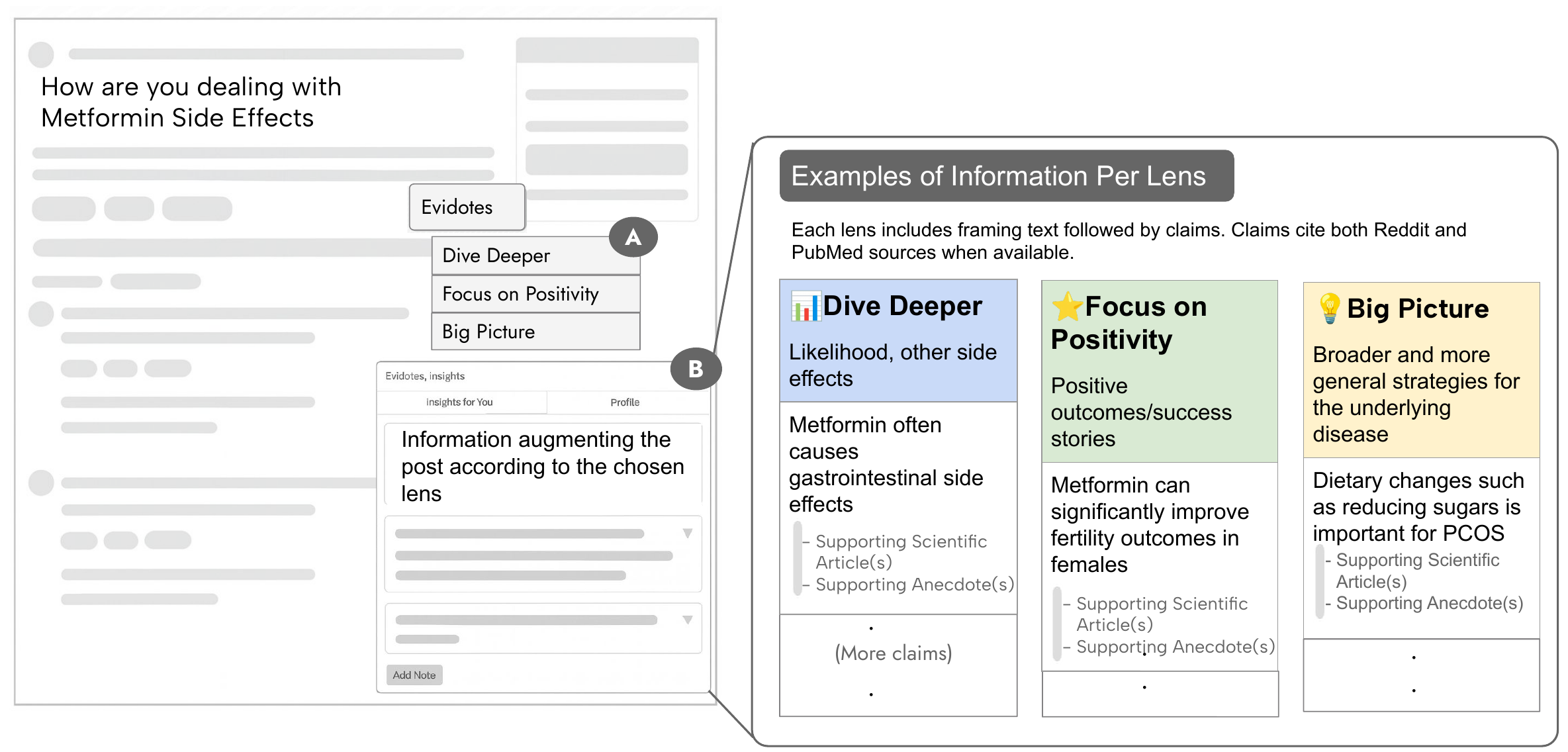}
  \caption{Evidotes supports patients in navigating health forums by combining scientific evidence with peer experiences. As a browser extension, it adds a button next to reddit posts and allows users to select different lenses (Dive Deeper, Focus on Positivity, and Big Picture) through a dropdown (A) based on their needs, and augments the original Reddit post with the synthesized information in an overlayed information pane (B) in form of claims. The figure shows a simplified view of Evidotes instantiation and the surfaced information across the three lenses.}
  \label{fig:teaser}
  \Description{A two-part figure illustrating the Evidotes browser extension and examples of its three evidence lenses. On the left, a Reddit post titled “How are you dealing with Metformin Side Effects” is shown with an Evidotes dropdown menu offering three modes: Dive Deeper, Focus on Positivity, and Big Picture. Below it, an embedded Evidotes panel displays additional information generated according to the selected lens. On the right, a zoomed-in panel shows example content for each lens. Dive Deeper includes likelihood and side-effect information with citations to scientific articles and anecdotes. Focus on Positivity highlights positive outcomes and success stories, also with citations. Big Picture provides broader strategies related to underlying conditions such as PCOS. Each lens section contains short example claims and placeholders indicating supporting sources.}
\end{figure*}

\section{Introduction}
Sarah, who is diagnosed with Polycystic Ovary Syndrome (PCOS), reads a Reddit post on the severe side effects of a medicine (Metformin) prescribed to her.
She spirals into a sea of worries: \emph{``What other side effects could happen to me?}'', \emph{``Should I consider an alternative treatment?}''
Sarah's experience exemplifies a fundamental paradox faced by millions navigating illness online: trying to seek comfort in peer support while being overwhelmed by new uncertainties.

Uncertainty is, in fact, an enduring feature of living with illness, shaping how individuals interpret symptoms, make treatment decisions, and envision their health trajectories~\cite{mishel1988uncertainty, Valeria2014, brown2020managing}.
Patients use online communities as a way to get access to lived experiences, emotional support, and practical insights from others facing similar health challenges~\cite{Gupta2018, Kingod2016, Coulson2018, Johansson2021}. Each shared experience represents a crucial data point in understanding the complex landscape of treatment outcomes and patient journeys. However, the very anecdotes that provide comfort and connection also generate new uncertainties for readers. 

Existing research on online health communities has often framed reader challenges on online health forums as an information quality problem alone - emphasizing credibility assessment, fact-checking, peer matching, and content filtering~\cite{burstin2023identifying, sylvia2020we, kanthawala2021credibility, chou2018addressing}. 
However, the core challenge is not solely quality, but also the inherent incompleteness and the emotions that these stories surface. Even the most trustworthy and relatable anecdotes often surface new questions and emotional concerns. Thus, users often go from post to post and comment to comment. 
Therefore, we need a system that provides scaffolds to help users navigate such uncertainties by providing complementary information, broader perspectives, and personally meaningful value.

Sarah's needs above extend beyond verification because the same Metformin that has helped others has also caused significant side effects in other cases.
Depending on her emotional state and information goals, she might need comprehensive research about Metformin alternatives, reassuring success stories to counter her anxiety, or a strategic overview of PCOS management options. 
Indeed, research consistently shows that people's needs differ when navigating medical uncertainty~\cite{mishel1988uncertainty, brashers2001communication}. 
Yet, existing systems offer little support for such contextualized needs that arise from engaging with peer posts, often leaving users in states of confusion and anxiety that drive counterproductive information-seeking behaviors.

In this paper, we present \textit{Evidotes} (Evidence + Anecdotes), a novel information support system, implemented as a Chrome Browser Extension, that augments health forum posts with additional scientific and anecdotal sources (Figure~\ref{fig:teaser}). 
We frame peer health posts as \emph{uncertainty triggers} that need contextualization rather than simple verification.
Accordingly, Evidotes shifts the emphasis from \emph{``Is this post accurate or relevant?''} to \emph{``What uncertainties has this post surfaced, and how might they be managed with additional information tailored to the user's needs?''}
Because users' informational and emotional needs vary widely, Evidotes provides scaffolding to guide discovery.
It allows users to choose how to augment a post by selecting whether to: (1) \textbf{Dive Deeper} into more similar information, (2) \textbf{Focus on Positivity} to get information focused on positive outcomes and success stories, or (3) Look at the \textbf{Big Picture} to help them zoom out and look at overall strategies for their disease management.
These options are presented as \emph{information lenses}, inspired by strategies patients often employ to manage health-related uncertainties, such as information seeking, information avoidance, emotional coping, and cognitive reappraisal~\cite{mishel1988uncertainty, brashers2001communication}.
To ground these design choices, we conducted a formative study with five participants.
Based on the findings, we implemented Evidotes as a browser extension (i.e., Chrome) that surfaces both anecdotal and scientific information—sourced via Reddit and PubMed APIs—within an information panel placed alongside the original post (Figure~\ref{fig:teaser}).

We evaluated Evidotes in a naturalistic study with chronic illness patients (\textit{N}=17). Across 130 lens interactions spanning 77 Reddit posts, participants reported significantly improved informational satisfaction (43\%, $p < .001$) and reduced emotional stress (44\%, $p < .001$) compared to their baseline self-reported usual browsing experiences. 
Through a mixed-methods analysis of user behaviors and interview transcripts, two key findings emerged.
Crucially, we discovered information symbiosis: rather than treating scientific and anecdotal sources as competing hierarchies, participants used each type to unlock the other's value. More specifically, anecdotal posts acted as an effective gateway to make scientific information more accessible (n=10/17), while also providing a \emph{complete picture} of the retrieved scientific information (n=6/17). Conversely, scientific research helped participants sift through the noise in online communities (n=7/17) while helping extract broader takeaways from anecdotal posts (n=5/17). This symbiosis challenges traditional hierarchical models and suggests that carefully blended approaches could better serve health information seekers navigating knowledge boundaries. Additionally, explicit lens choices helped users access diverse information without spiraling into anxiety (n=10/17) while discovering new uncertainty management approaches like emotional regulation (n=5/17). Together, these findings demonstrate how explicit user control over health information framing enables benefits unachievable through algorithmic inference alone.

These results demonstrated the effectiveness of Evidotes' design: blending anecdotes with scientific evidence at the level of individual claims and giving users agency through selectable lenses for augmenting Reddit posts. 
Through this work, we argue that health information systems should move beyond prioritizing accuracy and relevance alone. Instead, they should center user agency, acknowledge uncertainty as inherent to illness, and support effective integration of scientific and anecdotal information according to user needs.


\subsection{Contributions}
\begin{itemize}
    \item \textbf{Formative Study and Design}: 
    From a think-aloud study of patients browsing Reddit health forums (\textit{N}=5), we show how anecdotes act as uncertainty triggers, and draw four key design requirements for supporting users encountering such uncertainties.

    \item \textbf{Evidotes Artifact}: We present a novel information need support system, `Evidotes`, that augments peer health posts with synthesized and co-presented scientific documents and anecdotes retrieved through three different user-selectable information lenses. Evidotes is implemented as an open-source Chrome extension. The code for Evidotes is available at https://github.com/Sbali11/evidotes 

    \item \textbf{Empirical Validation and Mechanisms}: In a user study with chronic illness patients (\textit{N}=17, 130 lens interactions), Evidotes improved both informational satisfaction and emotional well-being. Interestingly, scientific and anecdotal sources created \textit{information symbiosis}: anecdotes made research accessible and contextual, while research helped filter and generalize from peer stories. Further, explicit lens choices helped users access diverse information without spiraling into anxiety and discover uncertainty management approaches like emotional management they had not previously considered. Based on these findings, we contribute design principles for health information systems that address diverse user needs through explicit user control and source integration, moving beyond algorithmic inference and filtering based on accuracy or relevance alone.

\end{itemize}

\section{Related Work} 
Our work builds on literature in uncertainty management, peer-driven health communication, and HCI systems for information support. We situate Evidotes within this literature to show how tools can help users not only access reliable information but also manage the uncertainty sparked by online health anecdotes.

\subsection{Uncertainty in Illness: Managing Health Unknowns}
Uncertainty in illness---both due to a lack of knowledge and the inherent unpredictability of disease---is widely recognized as an enduring aspect of illness~\cite{mishel1988uncertainty, brashers2001communication}. Rather than always aiming to eliminate or fully explain either kind of uncertainty, research in health communication emphasizes that management hinges on different strategies like information seeking, avoidance, emotional reframing, and social support~\cite{brashers2001communication, mishel1988uncertainty}. These strategies vary across illness types, user preferences, and stages of care. Similarly, communication guidelines for clinicians now emphasize emotional and cognitive reframing alongside factual clarity~\cite{simpkin2019communicating, Lichtstein2023, hart2021}. While not explicitly labeled as such, recent HCI research reveals how these strategies manifest in online health communities. Users, for instance, sometimes engage in ``hopeful avoidance'' where they selectively consume optimistic content in fertility forums, balancing emotional protection with information needs~\cite{shen2020analyses, DBLP:journals/pacmhci/KuA23}. Other documented digital behaviors include strategic information seeking, content avoidance to prevent emotional overwhelm, and seeking peer validation to reframe health concerns~\cite{brashers2001communication}.
However, current health community platforms provide minimal support for these uncertainty management strategies or exploration of what happens when explicit controls are given to users.

\subsection{Anecdotes as Catalysts that Surface Health Uncertainties} 
Online peer health anecdotes serve as critical ways to gather peer support, knowledge sharing, and collective sense-making ~\cite{neal2007online, solberg2014benefits, van2013using, mamykina2015collective}. For instance, Yang et al. explored how women with enigmatic conditions like vulvodynia use online communities to co-construct understanding of their poorly understood diagnoses, build individualized management plans, and share personal experiences that help others navigate medical uncertainty~\cite{DBLP:conf/chi/YoungM19}.

Some past work has applied the uncertainty management theory to offer one theoretical explanation for \textit{why} and \textit{how} individuals seek online health information~\cite{Rains2014, Zhuang2021, Miller2013, Delaney2021, Brashers2016}. While such research demonstrates that people seek peer experiences to manage existing illness uncertainty, we argue that health anecdotes paradoxically also create new uncertainties through a mechanism distinct from other information sources. We further report the findings of our formative study with chronic illness patients (\textit{N}=5), validating this.
Each anecdote provides a partial window into someone else's health trajectory, inherently incomplete yet vivid enough to prompt concern. When a patient reads about medication side effects they hadn't experienced, treatment failures in similar cases, or unexpected complications, they encounter what we term anecdote-triggered uncertainty - new concerns and information needs that emerge specifically from exposure to peer experiences. Through this reframing, we propose new design opportunities to help users manage the surfaced uncertainties.

\subsection{Current Technological Responses: Strengths and Limitations}

\subsubsection{Information Quality and Verification}
Much prior work on health information systems focuses on improving content quality and reducing misinformation. This includes developing automated classifiers for inaccurate or harmful posts~\cite{DBLP:journals/corr/abs-2407-07914, Bayani2023}, credibility assessment tools for end-users~\cite{kanthawala2021credibility, DBLP:conf/chi/HeuerG22, burstin2023identifying}, and moderation frameworks to uphold content norms~\cite{gatos2021hci, Skousen2020}. Recent efforts also identify high-similarity posters to personalize content~\cite{DBLP:journals/pacmhci/LevonianZNMTY25}. However, these approaches treat anecdotes as independent information objects to be verified or filtered without recognizing their interpretive impact. Even accurate stories can generate anxiety, doubt, or new questions requiring different system support. Addressing this challenge requires systems that support emotional and cognitive meaning-making, not just fact-checking.

\subsubsection{Anecdotes as Emotional and Social Infrastructure}
Health stories serve as affective resources, enabling validation, solidarity, and mutual support~\cite{van2008self, nakikj2017park, DBLP:conf/chi/ChenGHMS25}, with studies showing that peer sharing benefits both storytellers and readers~\cite{Yang20191, Yang20192}. However, emotionally charged narratives can propagate distress, as negative peer content can heighten user anxiety~\cite{DBLP:journals/epjds/SalatheVKH13, Park2017, Zhao2024} and create barriers to information consumption~\cite{DBLP:conf/chi/DixonABRL22}. Some past research has proposed addressing this through AI-driven emotional support interventions like therapeutic chatbots~\cite{Morris2018, Fitzpatrick2017} and automated responders~\cite{DBLP:journals/pacmhci/WangWTPFZYMW21}. However, recent HCI work critiques the "emotion regulation" paradigm for imposing rigid expectations about "proper" emotional expression in care contexts~\cite{DBLP:journals/corr/abs-2504-12614}, advocating instead for "emotion support" approaches that accommodate the full spectrum of emotional experiences without expectations for resolution. Crucially, these approaches treat emotional regulation as separate from information consumption, missing the opportunity to leverage information presentation itself as an emotion regulation mechanism. Evidotes addresses this gap by implementing user-controlled information augmentation as a primary mechanism for emotional support, recognizing that the manner in which health information is contextualized and presented can directly influence users' psychological states and uncertainty management capacity. Rather than seeking to regulate or modify emotional responses to health information, Evidotes validates users' emotional experiences while providing agency over how they engage with uncertainty-triggering content.

\section{Formative Study}
People managing health conditions seek both scientific evidence and peer experiences, yet little is known about how needs arise during active browsing.
Understanding what triggers confusion and what users require in real time is key to supportive design.
To inform our system design, we conducted think-aloud sessions with chronic illness patients while they browsed Reddit health communities.

\begin{table*}[t]
    \centering
    \caption{Demographics and health information seeking behaviors of formative study participants (n=5). Participants represented 3 ICD-11 disease categories and 5 disease conditions, which they searched for during the study. Three participants preferred scientific information, one preferred anecdotal information, and one preferred an equal mix. All participants were aged 25-44, with 2 seeking health information weekly, 2 rarely, and 1 monthly.}
    \label{tab:formative-participants}
    \begin{tabular}{c|p{3.4cm}|p{6.3cm}|p{3cm}|p{0.75cm}|p{2cm}}
        \toprule
        \textbf{PID} & \textbf{Condition Diagnosed With and Searched For} & \textbf{ICD-11 Category} & \textbf{Baseline Preference} & \textbf{Age} & \textbf{Health Search Frequency} \\
        \hline
        U1 & Polycystic Kidney Disease & Diseases of the genitourinary system & Mostly Scientific & 35-44 & Monthly \\
        U2 & ADHD & Mental, behavioral or neurodevelopmental disorders & Mostly Anecdotes & 25-34 & Weekly \\
        U3 & Uterine Fibroids & Diseases of the genitourinary system & Equal Mix & 25-34 & Rarely \\
        U4 & Insomnia & Mental, behavioral or neurodevelopmental disorders & Mostly Scientific & 35-44 & Rarely \\
        U5 & Migraine & Diseases of the nervous system & Mostly Scientific & 35-44 & Weekly \\
        \bottomrule
    \end{tabular}
    \Description{A table summarizing demographics and health information–seeking behaviors of five formative study participants (U1–U5). Columns include: participant ID, diagnosed condition searched for, ICD-11 category, baseline preference for scientific versus anecdotal information, age range, and frequency of health information searching. Participants represent conditions across genitourinary, mental and behavioural, and nervous system ICD-11 categories. Preferences vary: some prefer scientific information, one prefers anecdotes, and one prefers an equal mix. Ages range from 25–44. Health information search frequency ranges from weekly to monthly to rarely.}
\end{table*}

Five adults (U1–U5, ages 24–53) living with diagnosed long-term conditions were recruited via a local online health portal (Pitt+Me ~\footnote{https://pittplusme.org}).  All participants self-reported past use of online forums for health information as well as general experience using Reddit. Each participant was asked to use a subreddit related to their condition naturally while sharing their screen. Table \ref{tab:formative-participants} mentions their demographic information.
We prompted them to ``think aloud'' their information navigation process while browsing, selecting, and reading specific posts. 
Each session took approximately 35 minutes.
Participants received a \$20~USD Gift Card for their participation. 
All sessions were recorded and subsequently transcribed. 

\subsection{Findings}

We analyzed the transcripts using reflexive thematic analysis~\cite{Byrne2021} to identify recurring patterns and themes, iteratively coding and identifying themes from the data.

\subsubsection{The Integration Paradox: Seeking Synthesis while Maintaining Separation.}

In the pre-study questions, all five participants described a clear distinction between scientific and anecdotal evidence.
For example, U2 explained, \textit{``If I'm trying to figure out general information, I'll use a general information website. But... for specific things... I try to find a personal anecdote.''} However, in practice, participants frequently engaged in implicit integration across evidence types, expressing appreciation for posts that blended both. As U1 noted, \emph{``I'm interested when people put a link here... they might link to sources that I'm not familiar with...occasionally they might connect you to something that's scientific''}. Yet, we observed that even these participants, while acknowledging the existence of such links, did not actually go to the external sources. This indicates that the effort to bridge source types was too high without tool support. In fact, by observing the participants' interactions, we found that even participants who preferred scientific information (n=3/5) or valued both scientific and anecdotal sources equally (n=1/5) stayed within Reddit threads rather than consulting external validated sources. This suggests that platform design and search friction outweighed stated information preferences in actual browsing behavior.


\begin{designinsightbox}
\textbf{DR1: Enable low-effort evidence integration.} Tools should scaffold a way for users to engage with different evidence types (e.g., surfacing relevant studies when users encounter confusing anecdotes) rather than requiring manual synthesis. Users currently integrate evidence only when it's already presented together, suggesting that reducing the effort needed for cross-referencing can support richer sense-making.
\end{designinsightbox}

\subsubsection{Anecdotes as Uncertainty Triggers.}
    As anticipated, anecdotes consistently generated new questions and concerns for participants. When encountering symptom descriptions from other patients, U3 expressed uncertainty about interpretation: \emph{``I don't even know if I have a concrete answer for that... I don't know how I would react to it.''} U2 articulated frustration when an anecdote highlighted gaps in their medical knowledge: \emph{``Why are they seeing it? Because I have no idea, and my nephrologist could never tell me.''}. Additionally, U1 expressed displeasure at the negativity bias, \emph{``So most people have negative stuff.''} These responses illustrate anecdotes' dual nature - they provide valuable lived experiences while simultaneously highlighting the variability and uncertainty inherent in health outcomes, and raising questions and concerns for the readers.

\begin{designinsightbox}
\textbf{DR2: Support emerging informational needs.} Tools should help users address new questions that arise from encountering anecdotes, rather than leaving them with unresolved concerns.\\
\textbf{DR3: Address emotional impact.} Beyond information needs, systems should help users process the emotional distress that can arise from reading health anecdotes.
\end{designinsightbox}

\subsubsection{Serendipitous Discovery Competes with Focused Information Goals.} Most participants (n=4/5) relied on the Reddit feed rather than search. This unstructured browsing fostered serendipitous discovery, surfacing personally meaningful content they had not actively sought.
     For instance, U1 gravitated toward posts by \emph{``relatively young men''} with his condition, and appreciated religious coping references that aligned with his spirituality. Similarly, U5 prioritized posts that \emph{``make sense for my schedule and lifestyle.''}. However, this same unstructured exploration also created vulnerability to distraction. U4 captured the tension: \emph{``Next thing you know, I'm not even reading about insomnia... because somehow, another post will [pull me away].''}

\begin{designinsightbox}
\textbf{DR4: Balance discovery with focus.} Design should accommodate serendipitous browsing—which surfaces personally relevant content—while helping users avoid getting lost or distracted from their original goals.
\end{designinsightbox}

\subsection{Summary}

We found that patients navigating online health communities occupy a middle space: they demonstrate critical awareness (``grain of salt'' skepticism, source recognition) yet remain trapped in unresolved uncertainty. They question 
individual anecdotes but lack tools to follow up, while they recognize scientific evidence as more reliable, but often fail to access it. This pattern suggests that patients aren't na\"ive nor fully empowered. Instead, they are aware of their uncertainty yet lack effective means to address it. 
This motivates our approach: tools that bridge the gap between critical awareness and practical navigation, helping patients move from recognizing uncertainty to actively managing it. Below, we describe our proposed system based on the derived design recommendations (DR1-DR4) and prior work.

\section{Proposed System: Evidotes}


\begin{figure*}[t]
  \centering
  \includegraphics[width=0.98\textwidth]{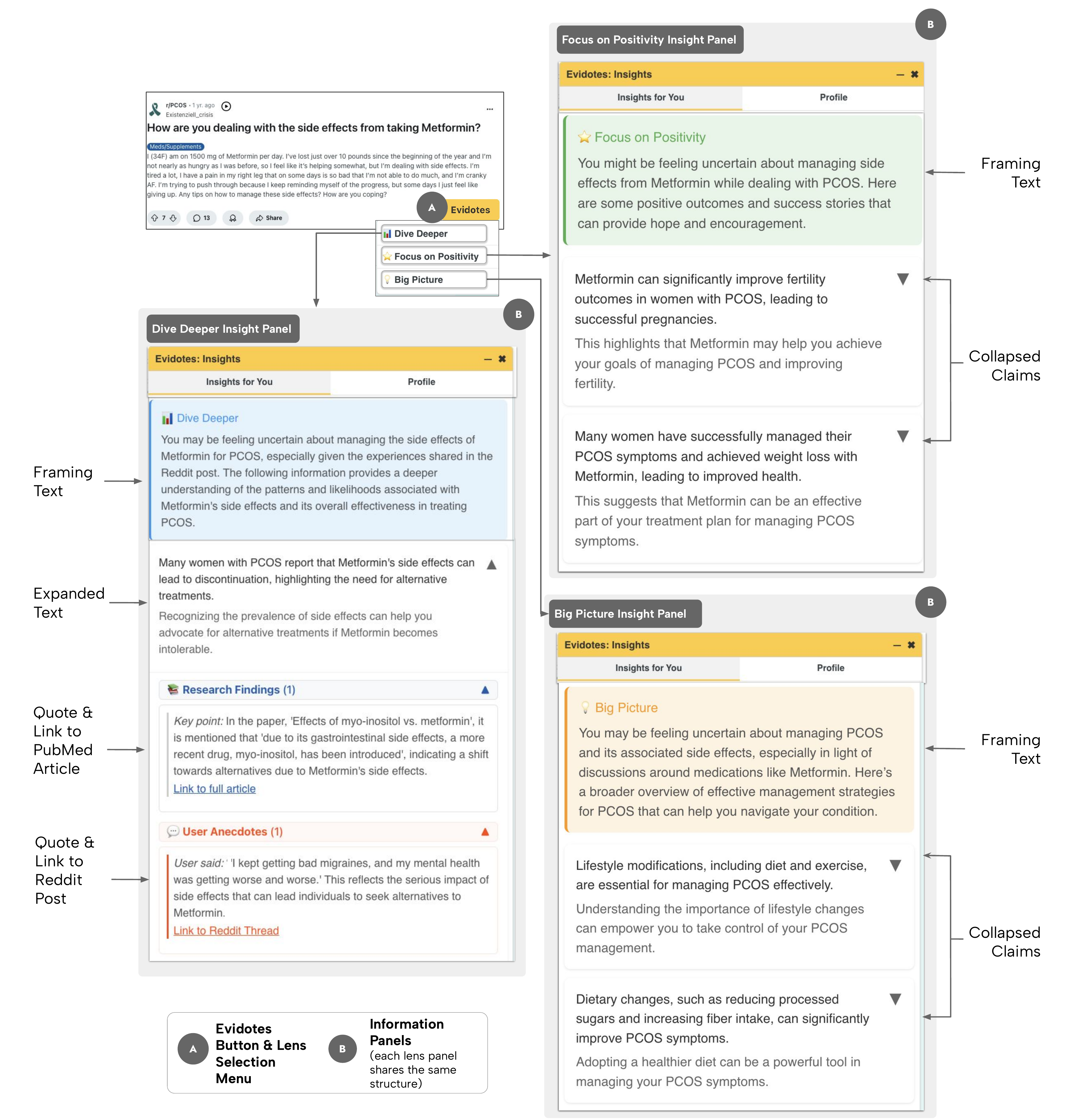}
  \caption{\textbf{Interface flow showing Evidotes' three lens options (A)  and resulting information panels (B) }. 1. Top Left: Original Post with Evidotes Button and Dropdown 2. Bottom Left: "Dive Deeper" panel with a claim expanded to show a detailed view with both Research Findings (scientific papers from PubMed) and User Anecdotes (Reddit posts) sections. 3. Top Right: "Focus on Positivity" panel displaying positive outcomes and success stories related to Metformin and PCOS. 4. Bottom Right: "Big Picture" panel providing a broader overview of PCOS management strategies, including lifestyle modifications and dietary changes. 
  }
  \label{fig:system-examples}
  \label{fig:system-overview}
  \Description{A multi-panel figure showing how the Evidotes extension presents three evidence lenses and their corresponding information panels. On the left, a Reddit post about Metformin side effects displays the Evidotes button and a dropdown with three options: Dive Deeper, Focus on Positivity, and Big Picture. Selecting a lens opens an information panel below the post. The “Dive Deeper” panel example shows framing text about uncertainty and an expanded claim explaining likelihood and side effects, followed by sections linking to scientific research findings (PubMed articles) and user anecdotes (Reddit posts). On the right, examples of the “Focus on Positivity” and “Big Picture” panels show their framing text and collapsed claims. The Focus on Positivity panel highlights positive outcomes, such as improved fertility or symptom management with Metformin. The Big Picture panel presents broader PCOS management strategies, including lifestyle modification and diet changes. Labels indicate panel structure, including framing text, expanded or collapsed claims, and citation links.}
\end{figure*}

We introduce Evidotes, a novel information needs support system that augments peer posts with additional findings from scientific articles and other peer posts retrieved through user-selectable lenses. Evidotes is implemented as a browser extension.
This section describes its key design, user interface, and information retrieval pipeline, along with our performance optimization strategy. 

\begin{sarahexample}

To illustrate system functionality, we use Sarah, a PCOS patient described in the Introduction, as a running example throughout this section.

\end{sarahexample}



    


\subsection{User Interface}

An overview of Evidotes is presented in~Figure~\ref{fig:system-overview}.
Based on the findings from our formative study, we consider each Reddit health post as a potential uncertainty trigger and implement support via the Evidotes extension interface. We implemented Evidotes as a browser extension to provide just-in-time support within users' natural browsing flows without requiring platform switching, and to enable easy generalization to other health platforms 
beyond Reddit.

\subsubsection{Entry Points and Lens Selection}
When users navigate to Reddit posts containing health-related content, the system automatically adds an ``Evidotes'' button positioned next to the post title. This placement signals the availability of additional evidence synthesis support. Clicking the button reveals a dropdown interface presenting three uncertainty navigation lenses inspired by convergent observations during the formative study and past related work. 
Our formative study revealed three distinct navigation patterns: participants wanted to (1) investigate specific concerns deeper, (2) manage emotional distress from negative posts, and (3) maintain strategic focus while exploring. These map broadly to established uncertainty management strategies: information seeking, emotional regulation, and cognitive reappraisal~\cite{mishel1988uncertainty, mishel1988finding} 
:

\begin{itemize}
    \item \textbf{Dive Deeper} retrieves additional context and evidence from both scientific articles and anecdotal Reddit posts about specific concerns mentioned in the original post. This lens addresses emergent information needs identified in the formative study where participants wanted a detailed exploration of specific concerns (DR2) and corresponds directly to the information-seeking strategy in prior literature~\cite{brashers2001communication}. For instance, if a post mentions a specific medicine, users can use this lens to retrieve more information about that medicine's side effects and benefits.
    \item \textbf{Focus on Positivity} retrieves related content that highlights uplifting insights from scientific research and positive Reddit anecdotes. This lens aims to help address the emotional impact that health posts and conditions cause (DR3) and supports emotional regulation strategies~\cite{mishel1988finding}. For instance, if an anxious reader reads a post talking about severe medication side effects or if Dive Deeper surfaces additional concerning cases, they can use this lens to surface recovery stories from people with similar conditions, and thus balance emotional protection with information needs.
    \item \textbf{Big Picture} returns other anecdotes and scientific papers highlighting broader patterns, alternative approaches, and general management strategies related to the health condition, helping users step back from post specifics and refocus on wider strategies. This lens facilitates serendipitous discovery by surfacing information across diverse management approaches for the condition (DR4). It also operationalizes cognitive reappraisal --- specifically, perspective broadening -- by recontextualizing immediate concerns within broader disease management frameworks~\cite{mishel1988uncertainty}. For instance, a reader focused on posts mentioning a specific medication could use this lens on one of the posts to discover complementary disease management strategies like diet changes for the same condition.
    \end{itemize}

These three lenses represent one instantiation of uncertainty-aware design grounded in both our formative study findings and established uncertainty management literature. While other strategies exist (e.g., information avoidance), these three represent actionable points where providing additional information can support navigation, aligning with our system's core mechanism of augmenting posts with contextual content. We chose these three lenses to balance meaningful diversity with cognitive simplicity.

\subsubsection{Synthesized Information Presentation}
After lens selection, Evidotes displays a draggable, minimizable panel (B) positioned at the bottom-right of the screen to avoid obscuring the original Reddit content.
The panel employs a tabbed interface with two sections: ``Insights for You'', which provides the evidence synthesis and serves as the focus of this research, and ``Profile'', which is limited to basic account settings (i.e. logout, view username). 

While retrieved content in ``Insights for You'' varies by lens, all three share a consistent presentation structure. Each view begins with contextual framing text that acknowledges the user's uncertainty and describes the type of information being presented (e.g., "You might be feeling uncertain about managing side effects from Metformin while dealing with PCOS. Here are some positive outcomes and success stories that can provide hope and encouragement"). This framing makes the lens's filtering approach transparent to users.
Below this, the interface displays synthesized claims (20-30 word summaries with key takeaways addressing the user's uncertainty). Each claim is supported by two expandable sections - Research Findings (PubMed articles) and User Anecdotes (Reddit posts) - containing direct links, relevant quotes, and context. This design addresses DR1 by reducing source-switching friction and enabling low-effort evidence integration. When a claim is supported by only one evidence type, we display the claim but leave the unsupported section empty, maintaining transparency about what evidence we could find while still validating the importance of the source that is present. 

\begin{sarahexample}

Sarah encounters a post from another PCOS patient describing excruciating leg pain that started two weeks after beginning Metformin. Having taken Metformin herself for six months, this triggers immediate uncertainty about her own treatment. Using Evidotes, she can select from three lenses: (1) Dive Deeper, (2) Focus on Positivity, or (3) Big Picture, each retrieving tailored information from both scientific articles (PubMed) and user stories (Reddit) displayed in a panel next to the original post. Figure \ref{fig:system-examples} shows what Sarah would see if she clicked on the different lenses. 

\end{sarahexample}

\subsection{LLM-Based Pipeline for Evidence Synthesis}
\label{sec:system-llm}

The backend pipeline implements our core design contribution: systematically translating psychological uncertainty management strategies into distinct retrieval and synthesis operations.
The pipeline is based on Retrieval-Augmented Generation (RAG) using Large Language Models (LLMs)~\cite{DBLP:journals/corr/abs-2312-10997}. To display relevant and useful information, we implement a three-stage RAG approach: lens-specific query generation, multi-source retrieval, and lens-specific synthesis.

First, Evidotes uses post content and the user lens selected to construct search queries tailored to both the health topic and the chosen uncertainty management strategy, using a query generation technique in RAG~\cite{DBLP:conf/acl/MaoHLSG0C20}. The system extracts health entities (conditions, treatments, symptoms) from the post and combines them with lens-specific framing. Dive Deeper queries focus on specifics and likelihoods (e.g., "PCOS Metformin side effects muscle pain likelihood"), Focus on Positivity targets success stories (e.g., "PCOS Metformin success stories"), and Big Picture queries broader management strategies (e.g., "PCOS management strategies guidelines"). This is expected to align evidence retrieval with both the health topic and the user's information needs.

 The system then performs parallel searches across two complementary source types, PubMed (for scientific articles) and Reddit (for anecdotal posts), for 10 results each, through asynchronous processing with a 25-second timeout to minimize response times. For PubMed, we use PubMed's Best Match algorithm~\cite{Fiorini2018} - a composite ranking that weights MeSH term matching, title/abstracts keyword frequency, and citation metrics. For Reddit, we use the Reddit search API sorting by 'relevance', leveraging Reddit's ranking that weighs query term matches, post scores, and comment engagement. All the retrieved results are passed to the synthesis stage without pre-filtering, allowing the LLM to assess relevance based on full context.

 Finally, we pass the retrieved sources to an LLM (GPT-4o-mini, temperature = 0) with a lens-specific synthesis prompt. The prompt instructs the LLM to: (1) tailor the response tone to the selected lens (analytical for Dive Deeper, hopeful for Focus on Positivity, or neutral for Big Picture), and (2) include direct quotes that support each claim. Each claim must cite at least one source (PubMed or Reddit), with preference for claims supported by both scientific and anecdotal evidence. Technical terms are paraphrased for lay readers. If no relevant sources are found, the system returns empty arrays rather than generating unsupported content. To further ensure reliability, all cited URLs are automatically checked for accessibility before inclusion. 
 We provide the full prompt in the supplemental files.

\begin{sarahexample}

Based on Sarah's lens selection, Evidotes analyzes the post content and generates tailored search queries for both PubMed and Reddit APIs:
(1) \textbf{Dive Deeper} → searches for ``PCOS Metformin side effects muscle pain likelihood'' and ``PCOS Metformin side effects timeline'' (2) \textbf{Focus on Positivity} → searches for ``PCOS Metformin success stories'' and ``PCOS recovery'' (3) \textbf{Big Picture} → searches for ``PCOS management'' and ``PCOS medicines''
The system then synthesizes retrieved sources into coherent claims organized under thematic headings. Sarah can view direct quotes from original sources and follow verification links for further exploration. 
Figure \ref{fig:system-examples} shows what Sarah would see if she clicked on the different lenses. 
\end{sarahexample}


\subsection{Implementation}

Evidotes operates through three integrated components working in concert to deliver contextual evidence synthesis. The frontend, implemented as a Chrome extension using Vue.js~\footnote{https://vuejs.org/}, provides the user interface and manages interaction with Reddit's existing page structure. The backend, built with Flask~\footnote{https://flask.palletsprojects.com/}, orchestrates evidence retrieval from external APIs and coordinates LLM-based synthesis using Agno-AI~\footnote{https://docs.agno.com/}, an open-source LLM orchestration framework that provides unified access to multiple language models. 
For the LLM, we used OpenAI's GPT-4o-mini. For PubMed, we use the NCBI E-utilities API (Entrez.esearch)~\footnote{https://www.ncbi.nlm.nih.gov/books/NBK25501/} for peer-reviewed research and Reddit's PRAW API~\footnote{https://praw.readthedocs.io/} for community discussions. The code for Evidotes is available at https://github.com/Sbali11/evidotes.


\section{User Study}

To examine how patients incorporated Evidotes into their illness navigation practices, we conducted a mixed-methods user study with real chronic illness patients. Each session was held over Zoom, recorded with consent, and lasted approximately 45 minutes. Participants received a \$20~USD gift card upon completion. Seventeen adults (P1–P17, ages 24–53) living with diagnosed long-term conditions were recruited via a local online portal for recruiting medical participants (Pitt+Me\footnote{https://pittplusme.org/}) and via online posts in social media groups (Reddit, X, LinkedIn). All participants self-reported past use of online forums for health information as well as general experience using Reddit. Table~\ref{tab:participants} presents their background information. Participants browsed subreddits related to their diagnosed conditions mentioned in the table.

Specifically, this  user study focused on these research questions

\begin{description}
\item[RQ1:] How do users interact with Evidotes?
\item[RQ2:] How does Evidotes affect users' information satisfaction and emotional outcomes when browsing health forums?
\item[RQ3:] How do users navigate scientific and anecdotal evidence when presented together in Evidotes?
\end{description}

\begin{figure*}[t]
  \centering
  \includegraphics[width=\textwidth]{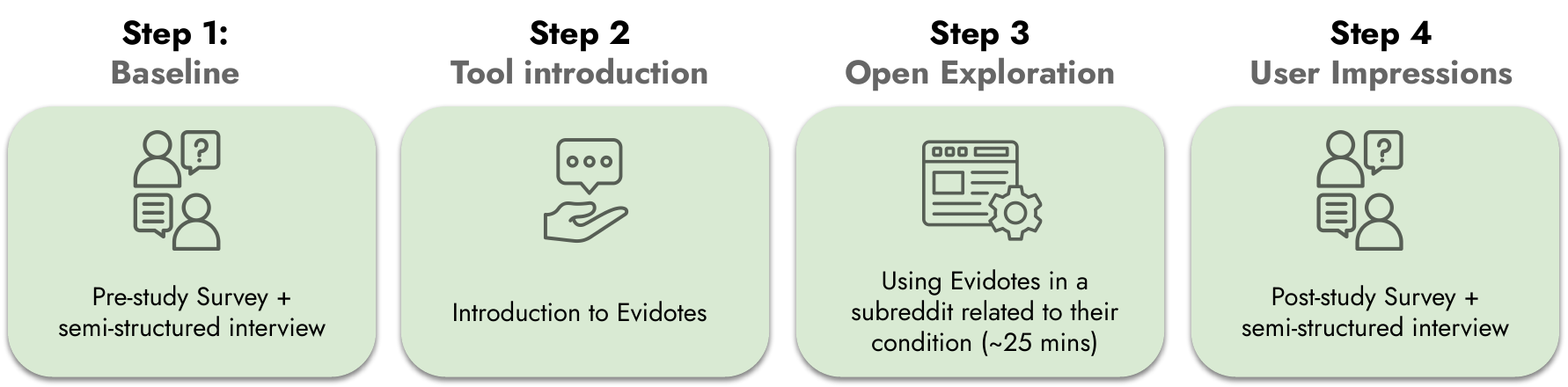}
  \caption{Overview of the procedure of our user study, where 17 patients used Evidotes for their own illness navigation. Icons from The Noun Project: Advice Icon by Alzam, Interview by Adrien Coquet, Browser by Muhammad Shabraiz}
  \label{fig:study-overview}
  \Description{A four-step diagram illustrating the procedure of the user study. Step 1, labeled “Baseline,” shows an icon of two people with a survey form and question mark, representing a pre-study survey and semi-structured interview. Step 2, “Tool Introduction,” includes an icon of a chat bubble and a hand, indicating an introduction to the Evidotes tool. Step 3, “Open Exploration,” shows a browser window with settings gear, representing participants using Evidotes on a subreddit related to their condition for about 25 minutes. Step 4, “User Impressions,” again shows two people with a survey form and question mark, representing a post-study survey and semi-structured interview.}
\end{figure*}

\subsection{Procedure}
The user study was conducted in four steps (Figure \ref{fig:study-overview}), described below.

\begin{table*}[t]
    \centering
    \caption{Demographics and health information seeking behaviors of study participants (n=17) with chronic conditions. Participants spanned 7 ICD-11 disease categories, with nervous system diseases most represented (n=5). Most participants preferred scientific information (n=8) or an equal mix (n=7), while 2 preferred anecdotal information. The majority were aged 25-44 (n=11) and sought health information weekly (n=9).}
    \label{tab:participants}
    \begin{tabular}{c|p{3.4cm}|p{6.3cm}|p{3cm}|p{0.75cm}|p{2cm}}
        \toprule
        \textbf{PID} & \textbf{Condition Diagnosed With and Searched For} & \textbf{ICD-11 Category} & \textbf{Baseline \newline 
        Preference} & \textbf{Age} & \textbf{Health Search \newline 
        Frequency} \\
        \hline
        P1 & Rheumatoid arthritis & Inflammatory arthropathies & Mostly Scientific & 25-34 & Weekly \\
        P2 & Diabetes & Endocrine, nutritional or metabolic diseases & Mostly Anecdotes & 35-44 & Weekly \\
        P3 & Breast Cancer & Neoplasms & Mostly Scientific & 35-44 & Weekly \\
        P4 & Arthritis & Diseases of the musculoskeletal system & Equal Mix & 45-54 & Weekly \\
        P5 & Asthma & Diseases of the respiratory system & Mostly Scientific & 45-54 & Daily \\
        P6 & Chronic Back Pain & Chronic pain & Mostly Scientific & 25-34 & Weekly \\
        P7 & Diabetes & Endocrine, nutritional or metabolic diseases & Mostly Anecdotes & 25-34 & Weekly \\
        P8 & Migraine & Diseases of the nervous system & Equal Mix & 35-44 & Daily \\
        P9 & Chronic Back Pain & Chronic pain & Mostly Scientific & 25-34 & Daily \\
        P10 & Migraine & Diseases of the nervous system & Equal Mix & 45-54 & Monthly \\
        P11 & Migraines & Diseases of the nervous system & Mostly Scientific & 18-24 & Monthly \\
        P12 & Diabetes & Endocrine, nutritional or metabolic diseases & Equal Mix & 55-64 & Monthly \\
        P13 & Migraines & Diseases of the nervous system & Equal Mix & 25-34 & Weekly \\
        P14 & Hypertension & Diseases of the circulatory system & Equal Mix & 35-44 & Weekly \\
        P15 & Migraines & Diseases of the nervous system & Mostly Scientific & 25-34 & Weekly \\
        P16 & PCOS & Diseases of the genitourinary system & Equal Mix & 18-24 & Daily \\
        P17 & Thyroid Issues & Endocrine, nutritional or metabolic diseases & Mostly Scientific & 25-34 & Rarely \\
        \bottomrule
    \end{tabular}
    \Description{This table presents data for 17 participants (P1-P17) across six columns. Column 1 shows participant IDs. Column 2 lists chronic conditions including migraines (5 participants), diabetes (3), chronic back pain (2), and single instances of rheumatoid arthritis, cancer, arthritis, asthma, hypertension, PCOS, and thyroid issues. Column 3 categorizes conditions using ICD-11 classifications, with nervous system diseases being most common (5 participants), followed by endocrine/metabolic diseases (4). Column 4 shows baseline information preferences: 8 participants preferred mostly scientific information, 7 preferred an equal mix, and 2 preferred mostly anecdotal information. Column 5 presents age ranges from 18-24 to 55-64, with most participants (11) falling in the 25-44 range. Column 6 indicates information-seeking frequency: 9 participants seek weekly, 4 daily, 3 monthly, and 1 rarely.}
\end{table*}


\begin{description}
    \item[Step 1: Baseline] Participants completed pre-study questionnaires about their typical past experiences with online health forums like Reddit before they read any posts in the session or were introduced to the tool. These baseline measures reflect their general experiences with health forum browsing without Evidotes. They then answered semi-structured questions about their health condition and general health information-seeking experiences.
    \item[Step 2: Tool Introduction] A researcher introduced them to the features of Evidotes and instructed them to install it on their computer/laptop.
    \item[Step 3: Open Exploration] After this, the participants engaged in a 25-minute session where they read self-selected posts related to the condition they were diagnosed with and used Evidotes whenever they wanted to. They shared their screen and were prompted to optionally ``think aloud'' while navigating the subreddit related to their health condition.
    \item[Step 4: User Impressions] Participants then completed the post-session measures and gave responses to a semi-structured interview.
\end{description}

\subsection{Measures}

To assess the impact of Evidotes, we measured participants' experiences across three key dimensions: information satisfaction, cognitive demand, and emotional well-being. We collected baseline measures reflecting participants' typical experiences with online health forums before introducing them to the tool, then repeated the same measures after participants used Evidotes. This within-subjects design enabled direct comparison of browsing experiences with and without tool support.
Table~\ref{tab:measures} presents the parallel questions administered at each time point. All questions used 5-point Likert scales ranging from 1 (Very Low) to 5 (Very High).

\begin{table*}[h]
\centering
\small
\caption{Pre-study and post-study measurement questions assessing information satisfaction, cognitive load, and emotional response.}

\label{tab:measures}
\begin{tabular}{p{8.9cm}|p{8.9cm}}
\toprule
\textbf{Pre-Study} \newline (Usual Health Information Seeking on Anecdotal Forums) & \textbf{Post-Study} \newline (With Evidotes) \\
\hline
How well do online anecdotal sources like Reddit meet your information needs? & 
How effectively did this session meet your information needs? \\
How mentally demanding is the task of interpreting information you read? & 
How mentally demanding was the task of interpreting the information you read? \\
How discouraged, stressed, or annoyed do you feel? & 
How discouraged, stressed, or annoyed did you feel? \\
(N/A) & The tool gave me information that I wanted to read/know\\
(N/A) & The tool made the experience less emotionally overwhelming\\
\bottomrule
\end{tabular}

\Description{A two-column table comparing pre-study and post-study measurement questions. The left column lists baseline questions about participants’ usual experiences: how well anecdotes meet information needs, how mentally demanding interpreting information is, and how discouraged or stressed they feel. The right column lists parallel post-study questions referring to the session using Evidotes, with two additional items asking whether the tool provided desired information and whether it made the experience less emotionally overwhelming.}
\end{table*}

\subsection{Explanation of User Study Design}
We designed our user study based on the following rationale.
Uncertainty triggers are idiosyncratic to each post-reader pairing; forcing participants to read \textit{identical} posts with and without Evidotes would remove the very context that gives rise to uncertainty because the second time users would have already seen a similar post before. Similarly, generic control tasks (e.g., ``browse with no tool for 10 min'') confound content variance with tool absence, and thus make it difficult to understand the true impact of the tool
We therefore adopted a naturalistic, within-session, mixed-methods approach that captures authentic trigger-response cycles.
Given the exploratory nature of this investigation and the complexity of authentic health information-seeking behaviors, we prioritized rich qualitative insights over statistical power. Our sample size (\textit{N}=17) aligns with our methodological choice and enabled us to identify behavioral patterns and design insights to answer the above research questions. Future work could employ longitudinal deployment studies comparing participants' health information practices before and after extended Evidotes use.

\section{Results}

The section introduces our findings drawn from three data sources: (1) in-tool usage logs tracking lens selections and user interactions, (2) verbatim transcripts from interviews and think-aloud sessions, and (3) pre-study and post-study quantitative responses. We analyzed the think-aloud and semi-structured interview questions using reflexive thematic analysis~\cite{Byrne2021} to identify recurring patterns and themes, iteratively coding and identifying themes from the data.

Overall, there were $\mathbf{293}$ total synthesized claims derived from the posts users interacted with. The synthesis process generated an average of $\mathbf{2.25}$ claims per post ($\pm 1.20$ SD). Each claim was supported by an average of $\mathbf{2.12}$ sources ($\pm 0.90$ SD), split nearly evenly between scientific and anecdotal evidence ($\mathbf{1.00}$ PubMed sources, $\pm 0.61$ SD; $\mathbf{1.12}$ Reddit sources, $\pm 0.57$ SD). A total of $\mathbf{37}$ claims were supported exclusively by PubMed sources, while $\mathbf{59}$ claims were supported exclusively by Reddit sources.

We first present technical validation of the accuracy of the information presented, i.e., the accuracy of the claims displayed by Evidotes. We then analyze user interaction patterns with Evidotes' three lenses (RQ1). Next, we present quantitative outcomes showing improved information satisfaction and reduced emotional stress from baseline browsing levels (RQ2). Finally, we examine the underlying mechanisms that enabled these improvements: how Evidotes facilitated symbiosis between scientific and anecdotal information, resolved the tension between discovery and overwhelm, and expanded participants' information-seeking strategies through cognitive scaffolding (RQ3).

\subsection{System Validation: Faithfulness of Citations and Topical Relevance}
\label{sec:results-accuracy}

\subsubsection{Faithfulness of the Generated Claims}
Evidotes generated 293 claims with 624 quote-source pairs across all posts participants viewed. The results below indicate that Evidotes maintains high fidelity in source attribution rather than fabricating or misattributing quoted material. To further mitigate the risk of LLM hallucinations~\cite{DBLP:journals/tois/HuangYMZFWCPFQL25}, Evidotes also provides users with direct source links so they can independently verify each citation.

\textbf{Computational Validation.}
To validate source attribution, we first applied spaCy-based\footnote{https://spacy.io/} text normalization to identify verbatim matches, and then computed BERTScore~\cite{DBLP:conf/iclr/ZhangKWWA20} to detect paraphrastic matches (using a threshold of $\ge$ 0.85, consistent with thresholds adopted in prior work in medical contexts~\cite{DBLP:journals/corr/abs-2505-00029, Shi2025} and beyond~\cite{Lowin2024}). Of these, 621 pairs were available for analysis (2 Reddit posts were deleted, and 1 was from a banned user). Overall, 96.7\% (601/621) of quotes were faithfully drawn from their cited sources: 77.3\% (480/621) appeared as direct matches, and an additional 19.5\% (121/621) met the high-similarity paraphrase threshold BERTScore. Each claim has multiple such sources. At the claim level, this translated to 89\% (275/293) of claims supported by all cited quotes, 11\% (15/293) were partially supported (i.e., at least one cited source provided a verbatim or high-similarity match), and only 3 claims lacked any supporting match. 

\textbf{Expert Validation.}
We recruited an independent medical doctor to evaluate responses, assessing 50 randomly selected faithful citations and all 20 flagged invalid citations. The evaluator judged responses for (1) medical accuracy, (2) potential harm, and (3) whether claims were validly derived from their quoted sources. 69/70 claims were medically accurate and none were potentially harmful. Only one claim was unverifiable: how "social gatherings often present challenges for people with diabetes". Among low-confidence citations (BERTScore<0.85), the evaluator confirmed 9/20 as invalid inferences but found 11/20 were correctly derived from sources, indicating valid paraphrasing. For high-confidence citations, all quote-source pairs were valid paraphrases. At the synthesized claim level, 43/50 claims were directly supported by their individual quoted sources. In 5/7 of the remaining claims synthesized information across multiple sources. For example, the claim that Xeloda \textit{"can stabilize disease progression, which is crucial for managing pancreatic cancer"} combined a Reddit post on treatment necessity with a PubMed article on pancreatic cancer aggressiveness—the "crucial" framing emerged from integrating both sources. While these cross-source syntheses could not be verified at the individual quote level, our validator confirmed them as medically accurate, demonstrating how complementary evidence can enhance claim value. The remaining two were more tenuous connections. For instance, in one of these cases the LLM used a Reddit discussion around a new implantable cell therapy for hypothyroidism treatment to infer \textit{Long-term management of hypothyroidism involves regular monitoring and adjustments in treatment to ensure optimal thyroid hormone levels.} The small number of such cases is promising, but this still indicates further scope in improving response quality.

\subsubsection{Topical Relevance}
Beyond faithful attribution, we validated whether retrieved information was relevant to the post and user needs.

\textbf{User Perceptions.}
The strongest evidence for topical relevance comes from participants themselves. 
Post-study measures showed high information satisfaction (4.6/5, up from 3.2/5 
baseline, p<.001), with 15/17 participants agreeing that "the tool gave me information I wanted to read/know." (additional details in Section \ref{sec:result-ux}). P13, who was one of the participants who gave a lower score, wanted information addressing multiple conditions mentioned in a single post. For example, when a post discussed both "PNES" and "Epileptic Seizures," the dive deeper feature only returned results for "PNES." While P13 acknowledged \emph{``I'm not saying that's not helpful,''} they suggested \emph{``something more helpful for this post alone would be able to find studies based on [both].''} 

\textbf{Computational Relevance Scores.}
To complement user perceptions, we computationally validated topical relevance across all retrieved claims. Using BERTopic~\cite{DBLP:journals/corr/abs-2203-05794} with GPT-4o-mini for semantically relevant representation, we identified 125 topics that clustered into $\mathbf{23}$ semantic groups (threshold $\approx 0.6$), yielding relevant interpretable themes like \textit{Managing Rheumatoid Arthritis Symptoms} and \textit{Long-Term Gabapentin Use Effects}. We assessed relevance through two hierarchical stages. First, \textit{Topic Intersection} validated whether a claim's topic matched the original post's topic, confirming $\mathbf{89.2\%}$ (554/621) of claims as relevant: 93.9\% for Dive Deeper, 87.8\% for Focus on Positivity, and 80.9\% for Big Picture. These differences align with design intent---Dive Deeper prioritizes thematic similarity while other lenses intentionally retrieve broader information. Second, \textit{Medical Condition Matching} used semantic similarity between subreddit and claim conditions, capturing an additional $\mathbf{9.0\%}$ (56/621) of claims. Overall, $\mathbf{98.2\%}$ (610/621) of claims were topically relevant: Dive Deeper ($\mathbf{99.4\%}$), Focus on Positivity ($\mathbf{99.4\%}$), and Big Picture ($\mathbf{94.7\%}$). This high relevance across all lenses indicates Evidotes maintained topical focus while broadening retrieval for emotional needs, strongly corroborating user-reported satisfaction.




\subsection{User Interaction Patterns (RQ1)}

\begin{figure*}[t]
  \centering
  \includegraphics[width=0.9\textwidth]{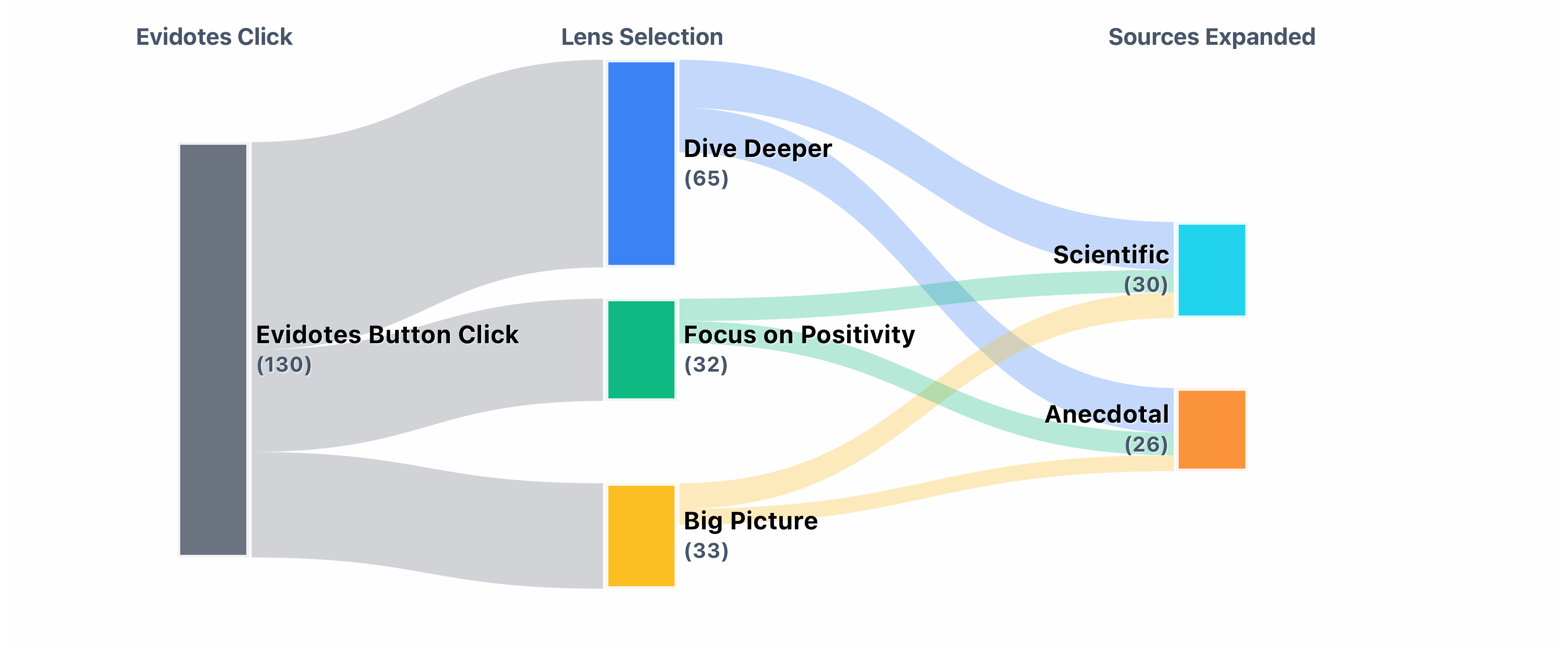}
  \caption{\textbf{User interaction trajectories showing lens selection and source expansion patterns. The Sankey diagram tracks flows from initial Evidotes clicks through lens selection, information consumption, and optional source expansion. "Sources Expanded" indicates clicks to view supporting materials—PubMed papers or Reddit posts. Flow thickness represents frequency. Key findings: Dive Deeper was most popular, ~50\% stopped at synthesized claims without expanding sources, and source expansion split roughly evenly between scientific and anecdotal materials. Users could view both source types per session; flows are separated for clarity.}
    }
  \Description{A Sankey diagram illustrating user interaction flows through the Evidotes interface. All paths begin with an “Evidotes Button Click” and split into three lens selections: Dive Deeper (65 clicks), Focus on Positivity (32 clicks), and Big Picture (33 clicks. Each lens further splits into two possible source expansions: viewing Scientific sources or viewing Anecdotal sources. Flow thickness corresponds to frequency; Dive Deeper shows the largest flow. Source expansion counts are roughly balanced, with 30 scientific expansions and 26 anecdotal expansions. Some flows end before source expansion, indicating users who stopped after reading synthesized claims.}
  \label{fig:user-trajectories}
\end{figure*}

The seventeen participants browsed condition-specific subreddits for an average of 25 minutes each. After the initial exploration, participants engaged with 77 posts total (\textit{M}=5.06, \textit{SD}=1.9 per participant) and triggered 130 Evidotes interactions (\textit{M}=9.1, \textit{SD}=3.2).
Figure~\ref{fig:user-trajectories} presents how each user interacted with different lenses and suggested sources in the study, showing diverse behavior with mixed usage for scientific and anecdotal sources. About half of the interactions with the information panel stopped at the overview of claims, while the other half participants interacted with both anecdotal and scientific information.

\subsubsection{Lens Selection and Switching Patterns}
\label{sec:lens-selection}

All participants switched between at least two lenses during their session. 
Overall, participants used the Dive Deeper lens most frequently, which accounted for about half the total number of lens interactions. Dive Deeper was the dominant initial choice (61/77 posts), with participants frequently progressing from Dive Deeper to Focus on Positivity (20 transitions) and from Focus on Positivity to Big Picture (17 transitions). No statistically significant relation between post tone and lens selection was found. However, the medical condition category showed a significant association with lens preference ($\chi^2=13.79$, $p=0.032$). Diseases of the nervous system and endocrine/nutritional/metabolic diseases, such as migraines and diabetes, showed a strong preference for Dive Deeper (51\% and 56\% respectively), while neoplasms favored Focus on Positivity (59\%). Diseases of the musculoskeletal system or connective tissue exhibited more balanced usage across all three lenses.  Given our sample size, these patterns could reflect individual differences, but they may also suggest that disease characteristics shape information needs - conditions with greater prognostic uncertainty or emotional burden (like cancer) may drive needs for positive framing and encouragement, while chronic conditions with complex symptom management (like diabetes, migraines, and PCOS) may benefit more from detailed, mechanistic information.

\subsubsection{Individual Behavioral Patterns}
While medical condition predicted overall lens preferences (Section \ref{sec:lens-selection}), individual usage patterns revealed how participants applied these lenses across different posts. We analyzed individual behavioral patterns using K-means clustering on features including lens usage percentages, switching frequency, and sequence diversity.
As a result, we found that the Dive Deeper usage percentage was the most discriminative factor, resulting in two clusters: \textit{diverse navigators} and \textit{focused deep-divers} (silhouette score = 0.3). 
The cluster analysis reveals that both participant characteristics and post-specific factors significantly influence lens selection, but they operate through different mechanisms. The 'diverse navigators' (15/17) adapted their lens choices across different posts based on contextual needs (entropy = 1.6). These participants employed different lens combinations depending on post content and momentary information needs. One participant illustrated this adaptive approach: \emph{``If I feel like I'm already overwhelmed, I'll go for Positivity.''} In contrast, the 'focused deep-divers' (2/17) maintained stable preferences across posts (76\% consistency, entropy = 0.9), focusing on only the 'Dive Deeper' lens. As P16 explained: \emph{``I read the post, and then, like I kind of just wanted posts that were like more related, like semantically related... I didn't really understand when I would use the Big Picture and the positive results options.''}
    
\begin{insightbox}
About 15/17 participants adopted the different lenses and changed the usage across different posts, indicating that most participants adopted the different lenses. The consistent lens-switching and overall usage patterns indicate a baseline level of acceptance, while also providing us with rich opportunities for further analysis. 
\end{insightbox}

\subsection{User Experience Measures (RQ2)}
\label{sec:result-ux}

\begin{figure*}[t]
  \centering
  \includegraphics[width=0.7\textwidth]{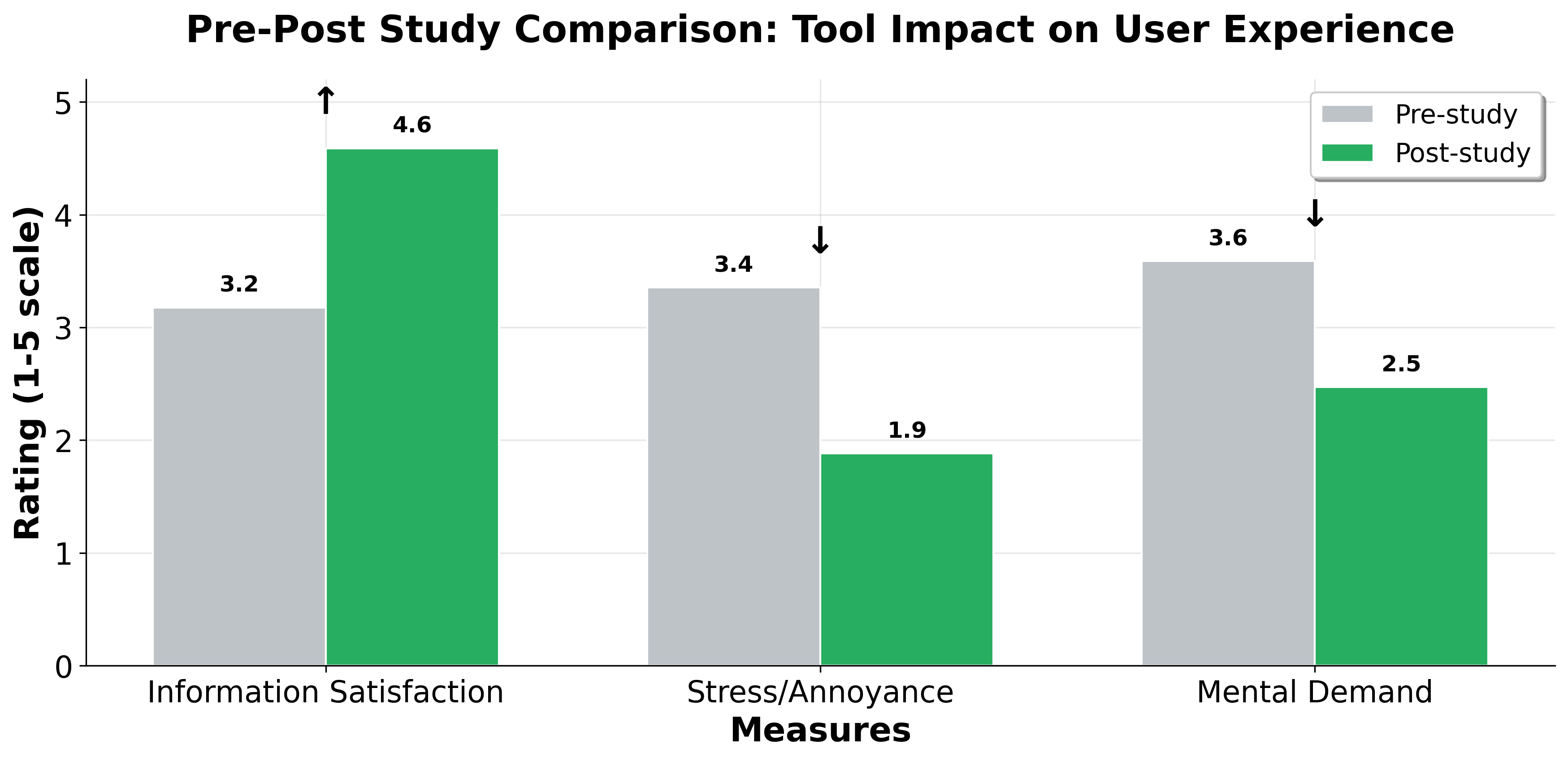}
  \caption{\textbf{Evidotes System Pre vs Post Study Results:} Participants reported significant improvements in information satisfaction (3.2 to 4.6, p<.001, d=1.49) and reduced stress (3.4 to 1.9, p<.001, d=1.23) after using Evidotes compared to baseline browsing. The graph shows before/after comparisons for key outcome measures. The arrows on top of bar groups show the ideal direction of change.}
  \label{fig:pre-post}
  \Description{A bar chart comparing pre-study and post-study ratings on three measures: Information Satisfaction, Stress/Annoyance, and Mental Demand. For Information Satisfaction, the pre-study rating is 3.2 and the post-study rating increases to 4.6. For Stress/Annoyance, the pre-study rating is 3.4 and the post-study rating drops to 1.9. For Mental Demand, the pre-study rating is 3.6 and the post-study rating decreases to 2.5. Upward or downward arrows above each measure indicate the desired direction of change. Overall, the chart shows increased satisfaction and reduced stress and cognitive load after using Evidotes.}
\end{figure*}


Figure~\ref{fig:pre-post} shows the results of the participants' responses to the pre-study and post-study questionnaires.
We confirmed that Evidotes' lens-based design achieved dual benefits: participants reported substantial improvements in both information satisfaction and emotional well-being.

First, information satisfaction increased substantially from 3.2 to 4.6 post-study (Change~=~43\%, $p<.001$, $d=1.49$). Nearly all participants (16/17) agreed the tool helped them perform tasks quickly, while the majority reported it supported them in achieving their goals (13/17) and left them confident about next steps (14/17). Strong agreement emerged that Evidotes made information easier to access (15/17) and provided relevant content (15/17).
P3 emphasized the system's broad applicability: \emph{``Anyone that likes to seek for health-related information online would be happy to reach this because it is specific and detailed enough for any user.''} 
Critically, Evidotes achieved this personalization without requiring health disclosures. Further, multiple participants (6/17) expressed a desire to use a tool like Evidotes for longer and on other platforms such as Instagram and Facebook. Some participants, (e.g., P5), however, were disappointed that Evidotes only supported text-based posts as they wanted to engage in some image-heavy posts. Further, while most found Evidotes streamlined information processing, some participants noted increased mental demand because of the new kind of retrieval. P8, for instance, mentioned that they found it \emph{``a bit mentally demanding... but for someone looking for something as important as health information, it's not a big deal.''}. This highlights the tension between informational richness and cognitive accessibility.

Secondly, participants reported significant reductions in stress and emotional burden (3.4 to 1.9, Change~=~44\%, $p<.001$, $d=1.23$), with 16/17 agreeing that Evidotes reduced emotional overwhelm. 
Several participants noted that Evidotes supported productive curiosity rather than anxiety. P5 described how positive framing helped: \emph{``The fact that it focuses on the positive... that counts for me. I'm like, it's not all bad. There is something good out of it.''}. Beyond explicit mood regulation, participants appreciated how the system provided structured pathways through overwhelming information landscapes.
\begin{insightbox}
These results validate the system design and suggest that Evidotes supported not just the information navigation but also emotional coping. Participants valued the system's flexibility, specifically the ability to choose different approaches based on momentary needs, while maintaining a consistent, learnable interface.
\end{insightbox}

\subsection{Evidotes Unlocked Information Symbiosis between Anecdotes and Scientific Evidence (RQ3)}


We further carefully analyzed the participants' responses in the exit interviews with thematic coding~\cite{Byrne2021} to understand how Evidotes helped them make sense of both scientific and anecdotal evidence.
In post-study interviews, 11/17 participants reported benefits from scientific-anecdotal integration beyond simply accessing more information, suggesting an ``information symbiosis.''
We found that this symbiosis manifested in two key ways: (1) enabling bidirectional navigation between scientific evidence and anecdotes, and (2) resolving the fundamental specificity-generalizability trade-off between the two sources. 

\subsubsection{Improved Navigation and Discovery Through Connection:}

Before using Evidotes, participants described persistent challenges navigating health information online. Almost all noted that while the internet contains abundant information, finding what they need remains difficult: scientific articles are hard to locate and understand, while Reddit forums contain valuable insights buried within substantial noise. Evidotes resolved this by creating bidirectional pathways between information types, where each source enhanced access to the other.\\

    \textbf{Anecdotal posts as effective gateways to relevant scientific information}\\ Many participants (10/17) appreciated how Evidotes enabled direct access to research from posts they were already reading. As P15 mentioned, \emph{``When I regularly search, it's hard to have to go out of your way to pull up the scientific. So it's just nice that it's provided for you.''}, Reddit posts acted as particularly effective gateways to relevant scientific information. They help "\textit{bring things back to your memory that you might have forgotten about"} (same participant, P15) and can then look up. 
    Interestingly, even participants with strong scientific preferences, like P16, found value in this anecdote-driven discovery approach: \emph{``I think it's a pretty good tool... like just having the scientific papers is like very helpful... I just wanna know what has been researched about this, like, how much do we know what have been the results.''}\\

    \textbf{Scientific information as a way to navigate peer stories}\\
    7/17 participants mentioned valuing scientific information as a way to understand whether they wanted to explore/use a Reddit post or not. For instance, talking about a specific post that didn't have scientific backing, P16 displayed critical evaluation: \emph{''I didn't find any like scientific information supporting it was just like people on Reddit were like, Oh, I made like these diet changes, and that helped with like this thing. But I didn't see any like scientific information supporting that. So then I'm not gonna...follow it until I see like evidence.''}.  P12 also appreciated the credibility granted by this combination, saying that while the anecdotes we presented were \emph{``a little more relatable, but then it combined that you also had the scientific side there to back up or balance them.''} This dual presentation thus enabled quick credibility assessments, helping participants efficiently filter signal from noise in Reddit discussions.

We conclude that the symbiotic relationship improved navigation in both directions: anecdotal information served as discovery pathways to relevant scientific research that participants would have otherwise missed, while scientific evidence helped participants more effectively navigate and filter Reddit discussions.

\subsubsection{Mutual Compensation for Specificity-Generalizability Tradeoffs:}

Anecdotes often feel too specific to others’ conditions, while scientific findings seem too generalized for personal concerns. As P11 noted: \emph{``[with anecdotes] you don't know what applies to you ... [while] scientific evidence ...is very cookie-cutter.''} Co-presenting both helped participants use the complementary strengths of each. After using Evidotes, P11 reflected: \emph{``a scientific evidence will give me the hard facts ... but Reddit thread can just help you expand into more...''}
Over half of the participants (9/17) echoed this view, though they varied in how they leveraged the integration.\\

    \textbf{Anecdotal stories provided contextual depth that made scientific findings personally applicable}\\
    Multiple participants (6/17) found that peer experiences enhanced the digestibility and personal relevance of scientific information. Anecdotes supplied rich contextual details and emotional authenticity that research abstracts typically lacked, creating what participants described as a ``complete picture.'' P10 explained the complementary value: \emph{``I liked it, because if I look at the scientific evidence and it really doesn't tell me too much then I can go to the user comments and they tend to give more answers than the scientific ones.''}
    Beyond informational completeness, anecdotes provided crucial emotional credibility that scientific evidence alone could not deliver. Participants described how peer experiences offered a \emph{``personal sense of relief''} (P12) and emotional validation that made abstract research findings feel trustworthy and applicable to their own circumstances.
     
     \textbf{Scientific evidence helped extract broader, generalizable takeaways from anecdotal posts}\\ 
    Conversely, several participants (5/17) found that the scientific context made it easier to identify relevant knowledge takeaways from individual stories. P9 explained \emph{``for scientific....it's more analyzed and like, it's giving the big picture of what what I'm looking for''}.   
    This enabled participants to extract useful insights from diverse anecdotal experiences without being overwhelmed by conflicting individual stories. As P1 noted, \emph{``And despite people having different experiences like, you can just know what to lean with, yeah. And it's reliable information.''}. 
    Furthermore, it helped participants contextualize extreme anecdotes within population patterns. P16 noted, \emph{``[With Reddit] you only get a very, very extreme subset of people like people who are like really negative or really positive and not like the median people. So then, I was looking at research papers, which was helpful from the tool.''}

\begin{insightbox}
    This section shows that, rather than maintaining artificial boundaries between scientific and experiential knowledge, co-presentation revealed them as complementary ways of knowing that enhance one another. These findings challenge the traditional information hierarchy lens of health information and information quality, suggesting an alternative definition: not the credibility of individual sources in isolation, but how different sources can be combined to meet users' information needs as a whole.
\end{insightbox}

\subsection{Emergent Effects: Interface Affordances for Managing Information and Emotion}

The above results confirmed the effectiveness of Evidotes in supporting patients' information navigation.
Finally, we discuss how different features on Evidotes worked synergistically to achieve it. 

\subsubsection{Bounded Exploration as a Means to Resolve the Discovery-Overwhelm Tension in Health Information}

Participants in our pre-study interviews mentioned a tension in online information seeking: the desire for discovery conflicted with the anxiety of infinite possibility. They described being \emph{``lost in the sea of information''}~(10/17). P14 captured this dilemma of how open-ended browsing \emph{``becomes harmful after a while''} and required self-imposed temporal boundaries-\emph{``stop after 30 minutes-because there's just so much information… I don't really know what to place value on.''} 
After using Evidotes, participants reported feeling reduced stress and anxiety as compared to their baseline browsing levels (Section~\ref{sec:result-ux}).
In their post-study survey, they mentioned that Evidotes helped reduce emotional overwhelm (16/17) while also helping them get the information they wanted (15/17). \\

\textbf{Sense of completeness that satisfied discovery needs without endless scrolling}\\ The retrieved information was deemed beneficial in helping receive relevant information while not getting overwhelmed. Participants repeatedly mentioned that Evidotes helped them get the \emph{``full story of the post''} (7/17), which supports the relevance and comprehensiveness of the synthesized claims. One commented,\emph{``otherwise you gotta like go through comments and comments and comments, and just scroll. This, just, you know, shows the most relevant comments to that to that post, or the most relevant scientific [information]''}. This sense of completeness was enhanced by co-presenting scientific and anecdotal information, as described in section 6.4.2. Further, for multiple participants (4/17), ``The Big Picture'' lens particularly exemplified strategic exploration by helping participants systematically discover previously unconsidered factors. P11 described how structured ``zooming out'' revealed new treatment approaches: \emph{``The big picture allows you to zoom out and see... could there be something else that I'm missing out... I wouldn't think of that if nobody had talked about sleep or posture.''} \\


\textbf{Lightweight anchoring that maintained focus during exploration}\\
    Some participants (3/17) also suggested that the presence of the Evidotes panel functioned as a lightweight anchor acting as a persistent reference point that prevented distraction. Appreciating the Evidotes structure P1 described, \emph{``[With Evidotes] You won't lose what you're looking for like sometimes, you know, sometimes you get distracted. And you're like, okay, what was I looking for? So it's being there. I feel like it's really important because you'll be able to, like, concentrate on what you're looking for.''} but \emph{``on every post that I click, it has a different deep dive. I can have like, maybe this medicine has this side effects, and this and then the other. One has 2 different other side effects, and it will have 2 different information. I love that actually.''}
    Similarly, P17, who ended up reading \emph{``about something to do with depression, something to do with mental health''} but appreciated how \emph{``I could get back easily, so that one was amazing...I'm able to get back to the main picture of what the topic I'm researching about is...''}.


\begin{insightbox}
We found that bounded exploration designed in Evidotes helped participants who wished for diverse information while reducing information overwhelm. 
By structuring information around user-selected lenses, the system helped participants satisfy their curiosity without losing focus or succumbing to information overload.
\end{insightbox}


\subsubsection{Focus on Positivity helped Reframe Emotion Regulation as an Information Strategy }

The Focus on Positivity lens revealed to participants—many of whom had previously limited themselves to purely analytical approaches—that emotion regulation can be a legitimate strategy for managing health uncertainty.
Multiple participants (5/17) discovered that actively seeking positive perspectives represented a valuable information strategy they had not previously considered.
P9's reflection captures this transformation: \emph{''It's rare to want to find a positive side… because you're really experiencing pain.''} P7 also described using positive accounts as \emph{``motivation to make my own success story''}. Most strikingly, rather than seeking only demographically similar experiences, which these participants had valued in the pre-study interview, participants began engaging aspirationally with dissimilar positive outcomes. Both P1 and P5 found value in treatment success stories despite their own experiences of side effects from the same treatment.
P1 mentioned that she used \emph{``dive deeper into it like, know more about it.''} After doing this, she decided on \emph{``focusing on the positive side of it, like how it has helped people."} Appreciating the results from Focus on Positivity, she generalized, saying \emph{``if you're searching for like medical stuff, and it dives deep.. sometimes you can be startled. At least it shows you like the positive side of it''}.  P5 echoed this: \emph{``The fact that it has that focus on the positive - that counts for me. It's not all bad. There is something good out of it.''} 

\begin{insightbox}
    We learned how explicit lens labels can function as cognitive scaffolds that legitimize previously unconsidered information strategies, expanding participants' approaches to uncertainty management beyond their default analytical frameworks, which participants appreciated.
\end{insightbox}

\section{Discussion}

Through this work and the user study, we discuss the broader implications of our work and potential future work for both healthcare and information systems.

\subsection{Design Implications}

\subsubsection{Reframing Health Information Systems as Uncertainty Navigation Tools}

The dominant paradigm in health information retrieval has been solely focused on getting more precise, filtered, and relevant information as a means to better support patients. However, as reconfirmed through our formative and user study, users don't seek information only to eliminate uncertainty - they seek tools to navigate it productively. Evidotes addresses this point by operating at the information layer, enabling contextual augmentation that addresses both informational and emotional needs. We contextualize peer stories by presenting scientific research alongside peer experiences, with user-selectable lenses guiding the retrieval and framing. This dual approach respects anecdotes' valuable functions while giving users evidence to critically evaluate individual stories. As Section 6.4 shows, this co-presentation creates \textit{"information symbiosis"} where each source type enhances the other's value. Its successful uptake by participants suggests that health information systems should optimize not just for accuracy or algorithmic relevance, but also for helping users navigate uncertainty. 

\subsubsection{Explicit Control for Health Information in an Era of Algorithmic Inference}
Contemporary health information platforms increasingly rely on algorithmic inference to predict user needs, for example, through personalized content recommendations~\cite{Adishesha2023Forecasting,DBLP:journals/corr/abs-2207-07915}.
Evidotes demonstrates an alternative approach: lightweight explicit user input that guides information synthesis through lens-based interactions.
By allowing users to choose their information lens, the system provides tailored results without requiring personal health disclosures.
Our user study demonstrated that this approach maintained strong information satisfaction while reducing emotional overwhelm.
This design aligns with prior literature on human-AI interaction that suggests providing a global control to users~\cite{DBLP:conf/chi/AmershiWVFNCSIB19}.
Moreover, interestingly, the lens selection process itself became a form of cognitive scaffolding.
Participants discovered uncertainty management strategies they had not previously considered, particularly using emotional regulation as a legitimate informational approach. This metacognitive benefit~\cite{DBLP:conf/chi/ChenYMHW25} - learning new ways to navigate health uncertainty - cannot be achieved through either algorithmic inference or traditional search, both of which assume users already know their information needs.

These findings suggest that patients' explicit input can play two roles: it gives algorithmic systems clear signals of user intent, and it also acts as a reflective tool that helps patients discover new strategies to handle uncertainty.
In healthcare contexts, where information choices carry serious consequences, this trade-off: accepting a small increase in cognitive effort in exchange for greater agency, may be especially worthwhile.
This approach thus charts a middle path between the passivity of algorithmic curation and the overwhelm of unguided exploration.

\subsection{Ethical Considerations: Balancing Agency, Accuracy, and Accessibility of Health Information}
\label{sec:disc-ethical}
Evidotes grapples with a fundamental tension: users regularly seek health information and support from peers on social networks that are noisy, often inaccurate, but still relatable. This creates persistent uncertainty for health information seekers. However, at the same time, social networks play a useful role for patients---peers speak their language and empathize, in ways doctors and expert-written sources often do not.  This is further complicated by patients' diverse, shifting needs, both informational and emotional, which we should support. We submit that such support, however useful, should not come at the cost of accuracy. Such systems should be \textit{both} flexible and accurate. 

Evidotes addresses these challenges by contextualizing individual health posts with additional peer anecdotes and scientific evidence, while providing users flexibility through user lenses. Additionally, our design prioritizes maintaining user awareness throughout the interface. For instance, when users select the Focus on Positivity lens, the framing text clearly states this filtering decision, ensuring they know they're seeing a curated subset. When claims lack scientific support, the empty scientific article section signals this gap while still presenting anecdotal evidence. This transparency enables users to critically evaluate unsupported anecdotal claims rather than unknowingly accepting potentially problematic information.

Despite these transparency mechanisms, we acknowledge scenarios where safeguards may prove insufficient: (1) users might not verify potentially hallucinated citations, and (2) chronic over-reliance on Focus on Positivity could create systematic optimism bias. Our study revealed that features enabling productive navigation---reduced effort, immediate accessibility, coherent integration---may lead to users not verifying underlying sources. This accessibility-verification tension represents a prevailing challenge as LLM-generated content becomes ubiquitous. While our validation showed acceptable accuracy (Section~\ref{sec:results-accuracy}), reduced verification suggests systemic risks as such tools scale. Future implementations should incorporate interventions that further promote critical engagement: progressive disclosure requiring source engagement before revealing claims~\cite{Muralidhar2025}, verification prompts for unverified information~\cite{DBLP:conf/chi/LeiserELKMSS24}, reflection-based comprehension checks, usage pattern detection with explicit signaling (\textit{``You've been focusing on positive outcomes: consider Dive Deeper for a complete picture''}), and longitudinal studies of downstream impact. Evidotes does not currently implement these because our exploratory study (n=17, single-session) was designed to validate the core design and understand emergent effects before adding adaptive restrictions.

\subsection{Limitations and Future Work}
While our measures of information satisfaction and emotional stress provide positive indicators for more effective uncertainty management, they serve as indirect proxies for uncertainty navigation behaviors. Future work should employ longitudinal deployment studies that directly assess users' uncertainty management behaviors over sustained uses, and how their uncertainty tolerance, management capabilities, and information seeking patterns change over time. 
The focused sample size (n=17) enabled rich pattern identification and qualitative insights that inform design principles, while future larger studies with the tool could establish broader statistical generalizability.
Additionally, our evaluation focused on Reddit health communities, and future work should examine tool effectiveness across other platforms (e.g., Facebook health groups, condition-specific forums) to understand how community characteristics influence uncertainty navigation behaviors. Given that some claims needed to be synthesized across sources (Section 6.1.1), future work could explore configurable synthesis parameters allowing users to choose between requiring each individual source to support returned claims versus permitting cross-source integration.

Beyond these individual and community-level considerations, disease management is inherently collaborative between patients and healthcare professionals.
We propose that Evidotes can act as a \textit{boundary object}—a flexible interface that bridges these groups by making users' uncertainty needs explicit and providing structured ways to navigate them.
Since clinical communication guidelines already emphasize tailoring information to patients' uncertainty preferences~\cite{mishel1988finding, mishel1988uncertainty, mishel1990reconceptualization}, such tools could help patients articulate both informational and emotional priorities more effectively during consultations.
In addition, if sharing Evidotes logs with clinicians becomes feasible, it may allow them to identify risks such as potential hallucinations or overreliance on a specific lens (as discussed in Section~\ref{sec:disc-ethical}), and to intervene appropriately to guide patients' information practices.
Future work should explore the privacy implications of such approaches and their influence on doctor–patient communication.

\section{Conclusion}
This paper reconceptualizes peer stories on health as inherently incomplete and as sources of uncertainty. We implemented and evaluated a novel tool, Evidotes, that augmented peer posts on Reddit with additional information from both anecdotes and scientific information across different user lenses.
Our user study (\textit{N}=17) confirmed that this interface increased participants' information satisfaction while simultaneously decreasing the emotional cost of searching for health information online as compared to their baseline levels.
Moreover, the study uncovered important user-driven information synthesis mechanisms. Based on the findings, we discussed future implications for designing health information systems that prioritize effective user-driven integration of scientific articles and anecdotes.

\begin{acks}
We thank Chinmay Kulkarni, Harivallabha Rangarajan, Prasoon Patidar, Beatriz Matos, and Anupama Sitaraman for their valuable feedback. 
This work was partially supported by NSF Award No. 2406099 and Carnegie Mellon University's Center for Machine Learning and Health.
The views and conclusions contained herein are those of the authors and should not be interpreted as representing the official policies or positions of the funding agencies.
\end{acks}

\bibliographystyle{ACM-Reference-Format}
\bibliography{sample-base}
\appendix
\newpage
\end{document}